\documentclass[aps,twocolumn,prb]{revtex4-2}
\usepackage{graphicx}
\usepackage{amsmath}

\draft 

\begin{document}

\title{Mode conversion and resonant absorption
in inhomogeneous materials with flat bands}

\author{Kihong Kim}
\email{khkim@ajou.ac.kr}
\affiliation{Department of Physics, Ajou University, Suwon 16499, Korea}
\author{Seulong Kim}
\affiliation{Department of Physics, Ajou University, Suwon 16499, Korea}

\begin{abstract}
Mode conversion of transverse electromagnetic waves into longitudinal oscillations
and the associated resonant absorption of wave energy in inhomogeneous plasmas is a phenomenon that has been studied extensively in plasma physics.
We show that precisely analogous phenomena occur generically in electronic and photonic systems where dispersionless flat bands
and dispersive bands coexist
in the presence of an inhomogeneous potential or medium parameter. We demonstrate that the systems described by the pseudospin-1 Dirac equation with two dispersive bands
and one flat band display mode conversion and resonant absorption in a very similar manner to $p$-polarized electromagnetic waves in an unmagnetized plasma
by calculating the mode conversion coefficient explicitly using the invariant imbedding method. We also show that a similar mode conversion process takes place in many other
systems with a flat band such as pseudospin-2 Dirac systems,
continuum models obtained for one-dimensional stub and sawtooth lattices, and two-dimensional electron systems with a quadratic band and a nearly flat band.
We discuss some experimental implications of mode conversion in flat-band materials and metamaterials.

\end{abstract}

\maketitle

\section{Introduction}

Recently, there has been growing interest in the electronic and photonic systems displaying
a dispersionless flat band in the band structure \cite{flach1,tang,im3}.
For the modes belonging to the flat band, the particle energy or wave frequency does not depend on the momentum or wave vector and the group velocity vanishes.
This has a strong effect on the behavior of quasiparticles and waves and can cause many interesting phenomena by greatly amplifying
the effects of various perturbations such as interactions and disorder \cite{liu,der,goda,shukla,ley,luck}.
Examples include superconductivity in magic-angle twisted bilayer graphene, flat-band ferromagnetism, and anomalous Landau levels \cite{cao,lieb,mielke1,mielke2,tasaki1,pons,balents,im1}.

Many models having one or more flat bands have been studied theoretically.
Two-dimensional (2D) lattices such as the Lieb, dice, and kagome lattices and one-dimensional (1D) lattices
such as the stub, sawtooth, and diamond lattices are among the examples \cite{dora,vic,flach,mizo}.
The low-energy physics of the aforementioned 2D lattices can be described by two Dirac cones intersected by a flat band
and modeled by the pseudospin-1 Dirac equation in 2D \cite{shen,urban,ocam,chan,kima}.
There also have been many recent attempts to realize flat-band systems experimentally \cite{vicen,muk,zong,bab,slot,xie}.

In this paper, we show that there exists another interesting phenomenon which has avoided
the attention of researchers until now, though it should occur generically in all flat-band systems unless forbidden by symmetry.
In plasma physics, the phenomenon termed (somewhat ambiguously) as mode conversion has been known for a long time and has played
a crucial role in explaining a variety of processes including the heating of solar corona and fusion plasmas and
the sudden appearance or disappearance of specific wave modes in space plasmas \cite{swan,mjo,hink,pp1,pp2,ehkim,ehkim2,yu1,yu2,jkps}. The simplest example is as follows.
In an inhomogeneous unmagnetized plasma where the plasma density $n$ varies smoothly along the $z$ direction,
the plasma frequency $\omega_p$ ($=\sqrt{4\pi ne^2/m}$), where $m$ and $e$ are the mass and charge of an electron, is also a function of $z$.
Let us consider a situation where a $p$-polarized electromagnetic (EM) wave of frequency $\omega$ is obliquely incident on this plasma and propagates within it.
If there exists a resonant region where $\omega$ is matched to the local plasma frequency, then the local dielectric permittivity
vanishes and the transverse wave excites
a longitudinal plasma oscillation there. Since the group velocity for the plasma oscillation mode is zero, the energy of the incident
wave is continuously converted into that of the plasma oscillation mode and is accumulated at the resonant region. Ultimately this energy will be dissipated as heat
and contribute to the heating of the plasma.

We point out that the plasma oscillation mode is an example of flat band. In an inhomogeneous plasma where this band crosses
the dispersive band describing EM waves, the energy can flow from the (fast) dispersive mode to the (slow) flat-band mode.
We will demonstrate that a precisely analogous phenomenon occurs in the systems described by the pseudospin-1 Dirac equation.
Furthermore, we will
show that similar phenomena take place in many other systems with flat bands
including pseudospin-2 Dirac systems, continuum models derived for 1D stub and sawtooth lattices, and a 2D model with a nearly flat band.

\section{Pseudospin-1 Dirac equation}
\label{sec_sp1}

\begin{figure}
\centering\includegraphics[width=6cm]{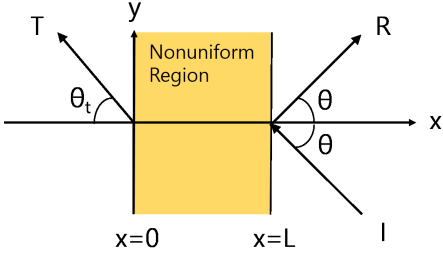}
\caption{Sketch of the configuration considered in Sec.~\ref{sec_sp1}. A plane wave is incident at an angle $\theta$ from the region
$x>L$ where $U=0$ onto the nonuniform region in $0\le x\le L$ where $U=U(x)$ and then transmitted at an angle $\theta_t$ to the uniform region $x<0$ where $U=U_t$.
If the wave is evanescent in the region $x<0$, then the transmittance $T$ vanishes and the angle $\theta_t$ is undefined.}
\label{fig_c1}
\end{figure}

The effective Hamiltonian that describes massive pseudospin-1 Dirac particles moving in the 2D $xy$ plane
in a 1D scalar potential $U=U(x)$ takes the form
\begin{eqnarray}
{\mathcal H}=v_F \left(S_x p_x+S_y p_y\right)+UI+M V,
\label{eq:ham1}
\end{eqnarray}
where $v_F$ is the Fermi velocity and $M$ ($=m{v_F}^2$) is the mass energy.
The $x$ and $y$ components, $S_x$ and $S_y$, of the pseudospin-1 operator are represented by
\begin{eqnarray}
S_x=\frac{1}{\sqrt{2}}\begin{pmatrix} 0& 1& 0\\ 1& 0& 1\\ 0& 1& 0\end{pmatrix},~~
S_y=\frac{1}{\sqrt{2}}\begin{pmatrix} 0& -i& 0\\ i& 0& -i\\ 0& i& 0\end{pmatrix}
\end{eqnarray}
and $I$ is the $3\times 3$ unity matrix.
The $x$ and $y$ components of the momentum operator, $p_x$ and $p_y$, are
\begin{eqnarray}
p_x=\frac{\hbar}{i}\frac{d}{dx},~~p_y=\hbar k_y,
\end{eqnarray}
where $k_y$ is the $y$ component of the wave vector.
We assume that the mass energy $M$ is a constant.
The mass term $MV$ describes the generation of the band gap between the conduction and valence bands and the position of the flat band.
For the matrix $V$, we choose
\begin{eqnarray}
V=\begin{pmatrix} 1& 0& 0\\ 0& -1& 0\\ 0& 0& 1\end{pmatrix}.
\end{eqnarray}
Then the flat band is located at the bottom of the conduction band if $M>0$
and at the top of the valence band if $M<0$ \cite{ocam}.
The size of the band gap is $2\vert M\vert$.

The time-independent Dirac equation in 2D for the three-component vector wave function $\psi$ [$=\left( \psi_1, \psi_2, \psi_3\right)^{\rm T}$] is
\begin{eqnarray}
{\mathcal H}\psi=E\psi,
\end{eqnarray}
where $E$ is the particle energy.
We can eliminate $\psi_1$ and $\psi_3$ using the equations
\begin{eqnarray}
\psi_1&=&-\frac{i}{\sqrt{2}}\frac{\hbar v_F}{E-M-U}\left(\frac{d}{dx}+k_y\right)\psi_2,\nonumber\\
\psi_3&=&-\frac{i}{\sqrt{2}}\frac{\hbar v_F}{E-M-U}\left(\frac{d}{dx}-k_y\right)\psi_2,
\label{eq:ff}
\end{eqnarray}
and obtain a single wave equation for $\psi_2$ of the form
\begin{eqnarray}
&&\frac{d}{dx}\left(\frac{\hbar v_F}{E-M-U}\frac{d\psi_2}{dx}\right)\nonumber\\&&~~~~+\left[\frac{E+M-U}{\hbar v_F}
-\frac{\hbar v_F{k_y}^2}{E-M-U}\right]\psi_2=0.
\label{eq:we1}
\end{eqnarray}
We assume that a plane wave described by $\psi_2$ is incident obliquely from the region $x>L$ where $U=0$
onto the nonuniform region in $0\le x\le L$ where $U=U(x)$ and then transmitted to the uniform region $x<0$ where $U=U_t$.
Then the wave number $k$ and the {\it negative} $x$ component of the wave vector, $p$, in the incident region
and the constant of motion $k_y$ are given by
\begin{eqnarray}
k=\frac{\sqrt{E^2-{M}^2}}{\hbar v_F},~~ p=k\cos\theta, ~~k_y=k\sin\theta,
\end{eqnarray}
where we assume that $E >M \ge 0$ and $\theta$ is the incident angle. A sketch of the configuration considered here is shown in Fig.~\ref{fig_c1}.

We introduce the dimensionless parameters $\epsilon$ and $\mu$ defined by
\begin{eqnarray}
\epsilon=1-\frac{U}{E-M},~~\mu=1-\frac{U}{E+M},
\end{eqnarray}
which are equal to each other in the massless case.
In the incident region, we have $\epsilon=\mu=1$.
In terms of the parameters $\epsilon$ and $\mu$, the wave equation, Eq.~(\ref{eq:we1}), can be written as
\begin{eqnarray}
\frac{d}{dx}\left(\frac{1}{\epsilon}\frac{d\psi_2}{dx}\right)+k^2\left(\mu-\frac{\sin^2\theta}{\epsilon}\right)\psi_2=0.
\label{eq:we0}
\end{eqnarray}
We notice that if we replace $\psi_2$, $\epsilon$, and $\mu$ with the $z$ component of the magnetic field $H_z$, the dielectric permittivity,
and the magnetic permeability, this equation has precisely the same form as the wave equation for $p$-polarized EM waves
propagating in the $xy$ plane. In Table \ref{tab:table1}, we make a comparison between the pseudospin-1 Dirac equation and the $p$ wave equation in a plasma.

\begin{table*}
	\caption{\label{tab:table1} Comparison between the pseudospin-1 Dirac equation and the $p$ wave equation in a plasma.}
	\begin{ruledtabular}
		\begin{tabular}{cccccc}
			& & $\epsilon$ & $\mu$ & flat band & local oscillation\\
			\hline
			& pseudospin-1 Dirac equation & $1-\frac{U}{E-M}$ & $1-\frac{U}{E+M}$ & $E=U+M$ & compact localized states \\
			& $p$ wave equation in a plasma & $1-\frac{{\omega_p}^2}{\omega^2}$ & 1 & $\omega=\omega_p$ & plasmon \\
		\end{tabular}
	\end{ruledtabular}
\end{table*}

We solve the wave equation in the presence of an arbitrary potential using
the invariant imbedding method \cite{kly,epl,sk1}. In this method, we first calculate the reflection and transmission coefficients $r$ and $t$ defined
by the wave functions in the incident and transmitted regions:
\begin{eqnarray}
\psi_2\left(x,L\right)=\left\{\begin{array}{ll}
  e^{ip\left(L-x\right)}+r(L)e^{ip\left(x-L\right)}, & x>L \\
  t(L)e^{-ip^\prime x}, & x<0
  \end{array},\right.
\end{eqnarray}
where $p^\prime$ is the negative $x$ component of the wave vector in the region $x<0$ and $r$ and $t$ are regarded as functions of $L$.
Following the procedure described in \cite{sk1}, we derive the exact differential equations for $r$ and $t$:
\begin{eqnarray}
&&\frac{1}{k}\frac{dr}{dl}=-\frac{i\cos\theta}{2}\epsilon\left(r-1\right)^2\nonumber\\&&~~~~~~~~~
+\frac{i}{2\cos\theta}\left(\mu-\frac{\sin^2\theta}{\epsilon}\right)\left(r+1\right)^2,\nonumber\\
&&\frac{1}{k}\frac{dt}{dl}=-\frac{i\cos\theta}{2}\epsilon\left(r-1\right)t\nonumber\\&&~~~~~~~~~
+\frac{i}{2\cos\theta}\left(\mu-\frac{\sin^2\theta}{\epsilon}\right)\left(r+1\right)t.
\label{eq:imb1}
\end{eqnarray}
For any functional form of $U$ and for any values of $kL$ and $\theta$,
we can integrate these equations from $l=0$ to $l=L$ using the initial conditions
\begin{eqnarray}
r(0)=\frac{\epsilon_2\cos\theta-\tilde p}{\epsilon_2\cos\theta+\tilde p},~~t(0)=\frac{2\epsilon_2\cos\theta}{\epsilon_2\cos\theta+\tilde p},
\end{eqnarray}
where
\begin{eqnarray}
&&\tilde p=\left\{\begin{matrix} \mbox{sgn}(\epsilon_2)\sqrt{\epsilon_2\mu_2-\sin^2\theta} &\mbox{if }\epsilon_2\mu_2\ge \sin^2\theta \\
i\sqrt{\sin^2\theta-\epsilon_2\mu_2} & \mbox{if }\epsilon_2\mu_2< \sin^2\theta \end{matrix}\right.,\nonumber\\
&&\epsilon_2=1-\frac{U_t}{E-M},~~\mu_2=1-\frac{U_t}{E+M},
\end{eqnarray}
and obtain $r(L)$ and $t(L)$.
The reflectance $R$ and the transmittance $T$ are obtained using
\begin{eqnarray}
R=\vert r\vert^2,~~T=\left\{\begin{matrix} \frac{\tilde p}{\epsilon_2\cos\theta}\vert t\vert^2
&\mbox{if }\epsilon_2\mu_2\ge \sin^2\theta \\ 0 & \mbox{if }\epsilon_2\mu_2< \sin^2\theta \end{matrix}\right..
\end{eqnarray}
In the absence of dissipation and mode conversion, the identity $R+T=1$ is satisfied.

The initial conditions $r(0)$ and $t(0)$ are the reflection and transmission coefficients for the case where there is no inhomogeneous layer (that is, $L=0$),
and therefore the incident region with $\epsilon=\mu=1$ and the transmitted region with $\epsilon=\epsilon_2$ and $\mu=\mu_2$ have a single interface at $l=0$.
They are derived from the continuity of $\psi_2$ and $\epsilon^{-1}d\psi_2/dx$
at the interface and are nothing but the well-known Fresnel coefficients. In the simplest case where the incident and transmitted regions have the same potential,
the initial conditions are trivially given by $r(0)=0$ and $t(0)=1$.

We point out that the invariant imbedding equations, Eq.~(\ref{eq:imb1}), become {\it singular} at the resonance point $x_r$ where $\epsilon=0$, which corresponds to $E=M+U(x_r)$.
This singularity causes mode conversion in a very similar manner to that
of transverse EM waves to longitudinal plasma oscillations in an inhomogeneous unmagnetized plasma. When a wave described by $\psi_2$ with finite group velocity
is incident obliquely on the inhomogeneous layer in $0\le x\le L$, it propagates up to the resonance point $x=x_r$, where the dispersive wave mode is strongly and resonantly coupled to the local flat-band state and the wave energy flows to the latter. Since
the group velocity associated with the flat-band state is zero, the energy is accumulated locally and is ultimately converted into heat.
In the steady state, a finite fraction of the energy of the incident wave is converted into that of the flat-band state.

In order to regularize the singularity,
we introduce a small imaginary part of $\epsilon$, $\epsilon_I$ ($>0$), in Eq.~(\ref{eq:imb1}) when calculating $r$ and $t$.
We find numerically that the absorptance $A$ ($=1-R-T$) converges to a finite value
in the limit $\epsilon_I\rightarrow 0$, if there exists a value of $x$ such that ${\rm Re}~\epsilon(x)=0$ in the region $0\le x\le L$.
We emphasize that this kind of absorption is not due to dissipation but due to the conversion of a propagating wave mode into a local
oscillating mode associated with the flat band. From now on, we will call $A$ as the mode conversion coefficient.

A clear signature of mode conversion is the occurrence of a singularity in the invariant imbedding equations, such as the $(\sin^2\theta)/\epsilon$ term in Eq.~(\ref{eq:imb1}).
In the systems with no flat band, there appears no singularity in those equations and mode conversion does not occur. In the case of the pseudospin-1/2
Dirac equation in the presence of inhomogeneous scalar and vector potentials, which describes single-layer graphene and does not have a flat band in its spectrum,
the invariant imbedding equations have been derived previously in \cite{sk11}. It has been verified that there appears no singularity and therefore no mode conversion.

\begin{figure}
\centering\includegraphics[width=8cm]{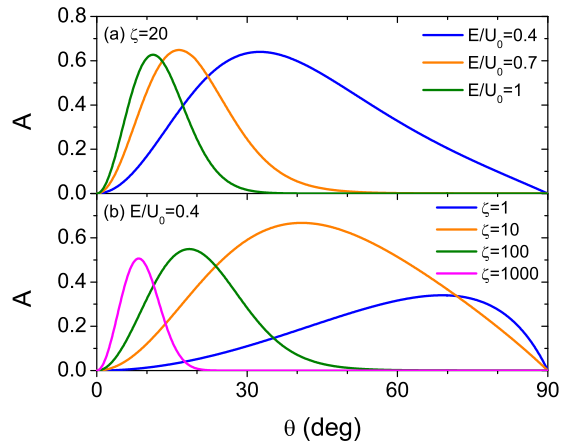}
\caption{Mode conversion coefficient $A$ obtained by solving Eq.~(\ref{eq:imb1}) for the configuration given by Eq.~(\ref{eq:slow}) plotted versus incident angle $\theta$,
(a) when $\zeta\equiv U_0L/(\hbar v_F)=20$, $M/U_0=0.2$, and $E/U_0=0.4$, 0.7, 1 and (b) when $E/U_0=0.4$, $M/U_0=0.2$, and $\zeta=1$, 10, 100, 1000.
In all calculations, $\epsilon_I$ is chosen to be $10^{-8}$.}
\label{fig1}
\end{figure}

\begin{figure}[h]
\centering\includegraphics[width=8.5cm]{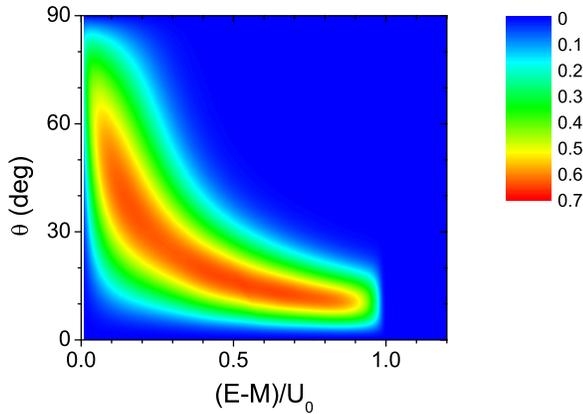}
\caption{Color graph of the mode conversion coefficient $A$ obtained by solving Eq.~(7) for the configuration given by Eq.~(8) as a function of $\theta$
and $(E-M)/U_0$, when $M/U_0=0.2$, $\zeta=20$, and $\epsilon_I=10^{-8}$.
$A$ vanishes for all $\theta$ if $(E-M)/U_0> 1$.}
\label{sfig1}
\end{figure}

To illustrate the mode conversion phenomenon, we consider a simple linear configuration of the potential
\begin{eqnarray}
  \frac{U(x)}{U_0}=\left\{ \begin{array}{ll}
  1,& \mbox{if } x<0\\
  1-\frac{x}{L},& \mbox{if } 0\le x\le L\\
  0, & \mbox{if } x>L
  \end{array} \right..
\label{eq:slow}
\end{eqnarray}
The resonance occurs inside the region $0\le x\le L$ if
the energy satisfies $M<E<U_0+M$.
In Fig.~\ref{fig1}(a), we plot $A$ versus $\theta$ when $M/U_0=0.2$ and $E/U_0=0.4$, 0.7, 1.
The thickness of the inhomogeneous slab satisfies $\zeta\equiv U_0L/(\hbar v_F)=20$. For the chosen values of the energy,
the resonance occurs within the inhomogeneous region and substantial absorption (that is, mode conversion) arises
in a wide range of the incident angle.

The efficiency of mode conversion is affected by the rate of the spatial change of the potential near the resonance point, which is measured by the parameter $\zeta$.
For a fixed value of $U_0$, $\zeta$ is proportional to $L$ and therefore is inversely proportional to the slope of the potential curve.
In Fig.~\ref{fig1}(b), we plot $A$ versus $\theta$ for the values of $\zeta$ equal to 1, 10, 100, and 1000, when $E/U_0=0.4$ and $M/U_0=0.2$.
We find that the mode conversion
becomes weaker for both small and large values of $\zeta$ and is strongest for some intermediate value of $\zeta$.

In Fig.~\ref{fig1}, we observe that $A$ always vanishes at $\theta=0$. This is because normally incident waves cannot couple to the local flat band mode in the present model.
In the equivalent case of mode conversion in an unmagnetized plasma, it has been well-known that normally incident transverse waves cannot excite longitudinal
plasma oscillations.
In order to provide a broader view of the mode conversion in pseudospin-1 systems, we show a color graph of the mode conversion coefficient as a function of the incident angle and the particle energy when $M/U_0=0.2$ and $\zeta=20$ in Fig.~\ref{sfig1}.
As it has been expected, when the energy satisfies $0<(E-M)/U_0<1$, there appears a wide range of the incident angle in which the mode conversion is substantially strong.

\begin{figure}
\centering\includegraphics[width=\linewidth]{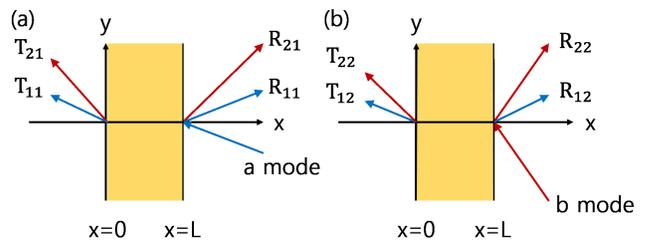}
\caption{Sketch of the configuration considered in Sec.~\ref{sec_sp2}. The medium described by the pseudospin-2 Dirac equation is birefringent, and in general
there exist two reflected waves and two transmitted waves for each of the incident $a$ and $b$ mode waves.
For the parameter values where some of the transmitted waves are evanescent, the corresponding transmittances vanish. When an $a$ mode wave is incident at an angle greater than $30^\circ$, the reflectance $R_{21}$ also vanishes. }
\label{fig_cf}
\end{figure}

\begin{figure}
\centering\includegraphics[width=8cm]{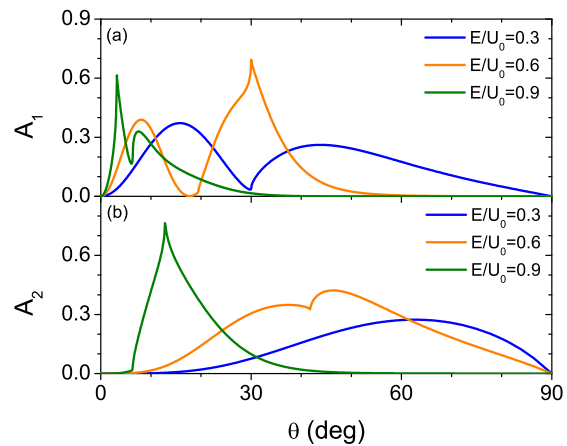}
\caption{Mode conversion coefficients (a) $A_1$ and (b) $A_2$ for massless pseudospin-2 Dirac particles in the configuration given by Eq.~(\ref{eq:slow}) plotted versus incident angle $\theta$,
when $\zeta=15$, $\epsilon_I=10^{-8}$, and $E/U_0=0.3$, 0.6, 0.9. $A_1$ ($A_2$) is obtained by calculating the absorptance when the incident wave is the $a$ ($b$) mode.}
\label{fig3}
\end{figure}

\begin{figure}[h]
\centering\includegraphics[width=8.5cm]{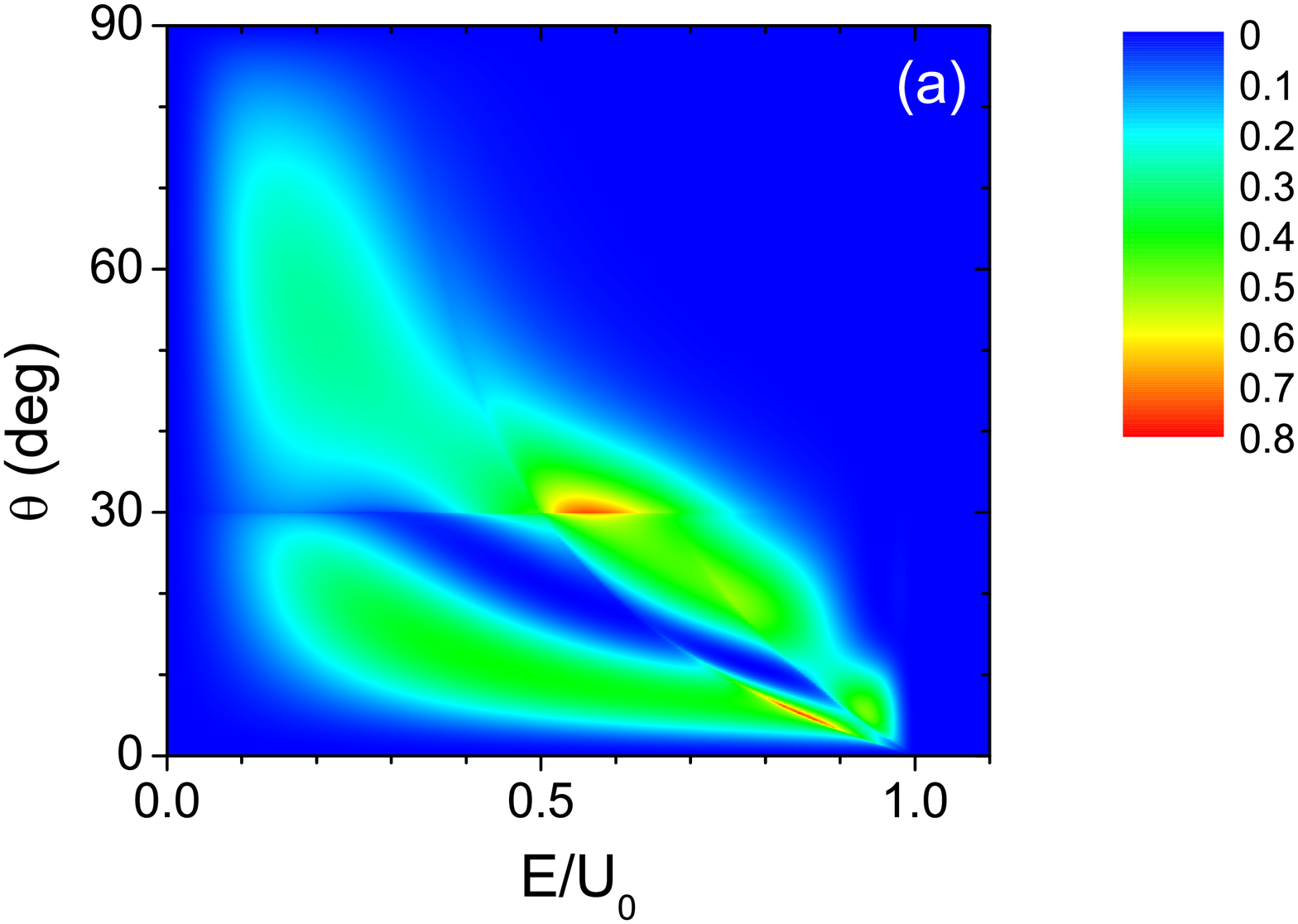} \includegraphics[width=8.5cm]{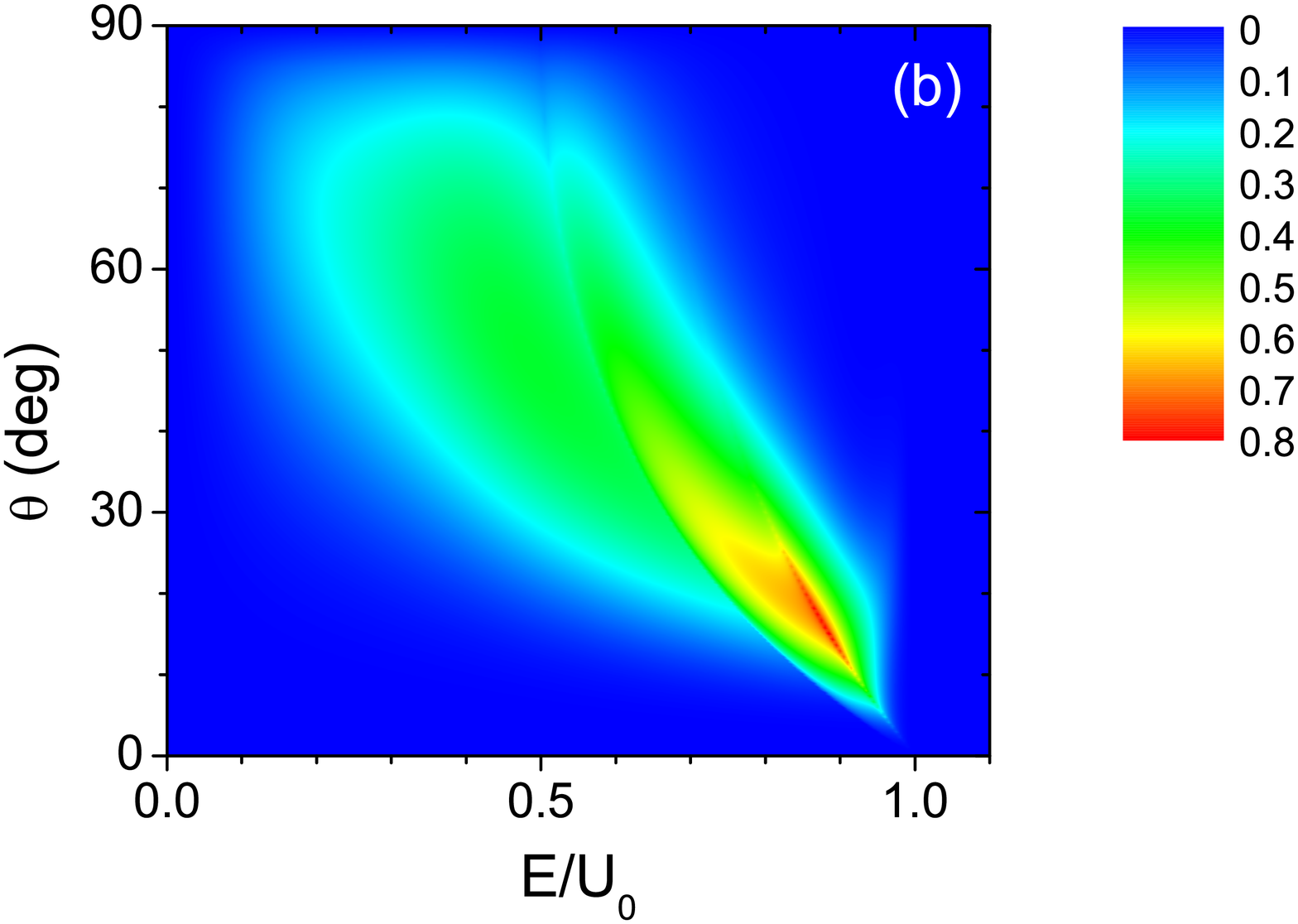}
\caption{Color graphs of the mode conversion coefficients (a) $A_1$ and (b) $A_2$ for massless pseudospin-2 Dirac particles as functions of $\theta$ and $E/U_0$, when $\zeta=15$ and $\epsilon_I=10^{-8}$. $A_1$ ($A_2$) is obtained by calculating the absorptance when the incident wave is the $a$ ($b$) mode.
Both $A_1$ and $A_2$ vanish for all $\theta$ if $E/U_0> 1$.}
\label{f1}
\end{figure}

\section{Pseudospin-2 Dirac equation}
\label{sec_sp2}

The band structure of pseudospin-$N$ Dirac systems with $N$ a positive integer consists of $2N$ dispersive
bands (that is, Dirac cones) and one flat band \cite{dora}. Therefore we expect all of these systems to display mode conversion.
We consider here the case of pseudospin-2 Dirac systems \cite{feng}.
The Hamiltonian that describes massless pseudospin-2 Dirac particles in 2D
in a 1D scalar potential $U=U(x)$ has a similar form as Eq.~(\ref{eq:ham1}), but with $M=0$ and
$S_x$ and $S_y$ given by
\begin{eqnarray}
&&S_x=\frac{1}{2}\begin{pmatrix} 0& 2& 0 &0 &0\\ 2& 0& \sqrt{6} &0 &0\\ 0& \sqrt{6}& 0 &\sqrt{6} &0\\ 0& 0& \sqrt{6} &0 &2\\ 0 &0 &0 &2 &0 \end{pmatrix},\nonumber\\
&&S_y=\frac{i}{2}\begin{pmatrix} 0& -2& 0 &0 &0\\ 2& 0& -\sqrt{6} &0 &0\\ 0& \sqrt{6}& 0 & -\sqrt{6} &0\\ 0& 0& \sqrt{6} &0 & -2\\ 0 &0 &0 &2 &0 \end{pmatrix}.
\end{eqnarray}

In the uniform case where the potential $U$ is constant, the eigenvalues of the Hamiltonian are given by
\begin{eqnarray}
&&E=U,\nonumber\\
&&E=U\pm \hbar v_F\sqrt{{k_x}^2+{k_y}^2},\nonumber\\
&&E=U\pm 2\hbar v_F\sqrt{{k_x}^2+{k_y}^2}.
\end{eqnarray}
Therefore the spectrum consists of two pairs of Dirac cones with different slopes which are intersected at the common apex by the flat band.
We now allow for the $x$ dependence of the potential $U$.
Starting from the pseudospin-2 Dirac equation for the five-component vector wave function $\psi$ [$=\left( \psi_1, \psi_2, \psi_3, \psi_4, \psi_5 \right)^{\rm T}$],
we can eliminate $\psi_1$, $\psi_3$,
and $\psi_5$ and derive two coupled wave equations for $\psi_2$ and $\psi_4$ of the form
\begin{eqnarray}
\frac{d}{dx}\left(A\frac{d\Psi}{dx} +B\Psi\right)+C\left(A\frac{d\Psi}{dx} +B\Psi\right)+D\Psi=0,\nonumber\\
\label{eq:s2}
\end{eqnarray}
where
\begin{eqnarray}
&& \Psi=\begin{pmatrix} \psi_2 \\ \psi_4 \end{pmatrix},~~A=\frac{1}{\epsilon}\begin{pmatrix} 1 & 0 \\ 0 & 1 \end{pmatrix},\nonumber\\ && B=\frac{k_y}{4\epsilon}\begin{pmatrix} 1 & 3 \\ -3 & -1 \end{pmatrix},~~
C=\frac{k_y}{8}\begin{pmatrix} 7 & 9\\ -9 & -7 \end{pmatrix},\nonumber\\ && D=\frac{\epsilon {k_0}^2}{8}\begin{pmatrix} 5 & -3 \\ -3 & 5 \end{pmatrix}+\frac{3{k_y}^2}{2\epsilon}\begin{pmatrix} -1 & 1 \\ 1 & -1 \end{pmatrix},\nonumber\\
&&\epsilon=1-\frac{U}{E},~~k_0=\frac{E}{\hbar v_F}.
\label{eq:abc}
\end{eqnarray}
In the uniform region where $U$ is constant, there are four solutions for the $x$ component of the wave vector, $p$, obtained from Eq.~(\ref{eq:s2}), which are
\begin{eqnarray}
p=\pm\sqrt{{k_0}^2\epsilon^2-{k_y}^2},~~p=\pm\sqrt{\frac{1}{4}{k_0}^2\epsilon^2-{k_y}^2}.
\end{eqnarray}
The $\pm$ signs represent the direction of the phase velocity.
We notice that there are two orthogonal eigenmodes obtained as linear combinations of $\psi_2$ and $\psi_4$, which are associated with the inner and outer cones and called
here as $a$ and $b$ modes respectively. These modes are characterized by different effective refractive indices
$\epsilon$ and $\epsilon/2$ and the system is birefringent. The $a$ and $b$ modes can alternatively be called as $h=1$ and $h=2$ modes respectively, where $h$ refers to the helicity.
The effect of the flat band is absorbed into the coefficients of Eq.~(\ref{eq:s2}).
In inhomogeneous media, $a$ and $b$ modes
interact with each other and with the local flat band mode.

In the uniform region, we can show that there exists a linear proportionality relation between $\psi_2$ and $\psi_4$, which is different for $a$ and $b$ modes and for the left-moving
and right-moving waves.
We obtain
\begin{eqnarray}
&&{\psi_4}^{(la)}=\eta_{la}{\psi_2}^{(la)},~~{\psi_4}^{(lb)}=\eta_{lb}{\psi_2}^{(lb)},\nonumber\\&& {\psi_4}^{(ra)}=\eta_{ra}{\psi_2}^{(ra)},~~{\psi_4}^{(rb)}=\eta_{rb}{\psi_2}^{(rb)},
\end{eqnarray}
where
\begin{eqnarray}
&&\eta_{la}=-\frac{{k_0}^2\epsilon^2}{\left(-\sqrt{{k_0}^2\epsilon^2-{k_y}^2}-ik_y\right)^2},\nonumber\\
&&\eta_{lb}=\frac{\frac{1}{4}{k_0}^2\epsilon^2}{\left(-\sqrt{\frac{1}{4}{k_0}^2\epsilon^2-{k_y}^2}-ik_y\right)^2},
\nonumber\\ && \eta_{ra}=-\frac{{k_0}^2\epsilon^2}{\left(\sqrt{{k_0}^2\epsilon^2-{k_y}^2}-ik_y\right)^2},\nonumber\\
&&\eta_{rb}=\frac{\frac{1}{4}{k_0}^2\epsilon^2}{\left(\sqrt{\frac{1}{4}{k_0}^2\epsilon^2-{k_y}^2}-ik_y\right)^2}.
\label{eq:eta}
\end{eqnarray}
The wave function is expanded in terms of $a$ and $b$ modes as
\begin{eqnarray}
\Psi=\begin{pmatrix} \psi_2 \\ \psi_4 \end{pmatrix}&=&\begin{pmatrix} {\psi_2}^{(la)}+{\psi_2}^{(lb)}+{\psi_2}^{(ra)}+{\psi_2}^{(rb)}
\\ {\psi_4}^{(la)}+{\psi_4}^{(lb)}+{\psi_4}^{(ra)}+{\psi_4}^{(rb)} \end{pmatrix}\nonumber\\&=&N_l \begin{pmatrix} {\psi_2}^{(la)} \\ {\psi_2}^{(lb)} \end{pmatrix}
+N_r \begin{pmatrix} {\psi_2}^{(ra)} \\ {\psi_2}^{(rb)} \end{pmatrix},
\end{eqnarray}
where
\begin{eqnarray}
N_l=\begin{pmatrix} 1 & 1 \\ \eta_{la} & \eta_{lb} \end{pmatrix},~~N_r=\begin{pmatrix} 1 & 1 \\ \eta_{ra} & \eta_{rb} \end{pmatrix}.
\label{eq:cf}
\end{eqnarray}
Since there are two propagating wave modes, we need to define the reflection and transmission coefficients $r$ and $t$ as $2\times 2$ matrices.
In our notation, $r_{21}$ ($r_{11}$) is the reflection coefficient when the incident wave is $a$ mode and the reflected wave is $b$ ($a$) mode.
Similarly, $r_{12}$ ($r_{22}$) is the reflection coefficient when the incident wave is $b$ mode and the reflected wave is $a$ ($b$) mode.
Similar definitions are applied to the transmission coefficients.

Following the procedure given in \cite{sk1}, we derive the invariant imbedding equations for $r$ and $t$:
\begin{widetext}
\begin{eqnarray}
&&\frac{dr}{dl}={N_{ri}}^{-1}\left\{-A^{-1}B+i\left(N_{li}+N_{ri}r\right)\left(A_iN_{li}P_i+A_iN_{ri}P_i{N_{ri}}^{-1}N_{li}\right)^{-1}
\left[-\left(iA_iN_{ri}P_i{N_{ri}}^{-1}+B_i\right)A^{-1}B+D\right]\right\}\nonumber\\
&&~~~~~~~~~\times\left(N_{li}+N_{ri}r\right)\nonumber\\
&&~~~~~~+{N_{ri}}^{-1}\left\{A^{-1}+i\left(N_{li}+N_{ri}r\right)\left(A_iN_{li}P_i+A_iN_{ri}P_i{N_{ri}}^{-1}N_{li}\right)^{-1}
\left[\left(iA_iN_{ri}P_i{N_{ri}}^{-1}+B_i\right)A^{-1}+C\right]\right\}\nonumber\\
&&~~~~~~~~~\times\left[iA_i\left(-N_{li}P_i+N_{ri}P_ir\right)+B_i\left(N_{li}+N_{ri}r\right)\right],\nonumber\\
&&\frac{dt}{dl}=it\left(A_i N_{li}P_i+A_i N_{ri}P_i{N_{ri}}^{-1}N_{li}\right)^{-1}\Big\{\left[-\left(iA_i N_{ri}P_i{N_{ri}}^{-1}+B_i\right)A^{-1}B+D\right]\left(N_{li}+N_{ri}r\right)\nonumber\\
&&~~~~~~+\left[\left(iA_iN_{ri}P_i{N_{ri}}^{-1}+B_i\right)A^{-1}+C\right]\left[iA_i\left(-N_{li}P_i+N_{ri}P_ir\right)+B_i\left(N_{li}+N_{ri}r\right)\right]\Big\},
\label{eq:imb2}
\end{eqnarray}
\end{widetext}
where $A_i$, $B_i$, $N_{li}$, and $N_{ri}$ are the values of $A$, $B$, $N_l$, and $N_r$ in the incident region obtained by setting $\epsilon=\epsilon_i=1$ in Eqs.~(\ref{eq:abc}) and (\ref{eq:eta}).
These equations are integrated using the initial conditions of the form
\begin{widetext}
\begin{eqnarray}
&&r(0)=\left(A_i N_{ri}P_i+A_t N_{lt}P_t{N_{lt}}^{-1}N_{ri}-i B_i N_{ri}+iB_t N_{ri}\right)^{-1}
\left(A_i N_{li}P_i+A_i N_{ri}P_i{N_{ri}}^{-1}N_{li}\right)-{N_{ri}}^{-1}N_{li},\nonumber\\
&&t(0)=\left(A_i N_{ri}P_i{N_{ri}}^{-1}N_{lt}+ A_t N_{lt}P_t-i B_i N_{lt}+i B_t N_{lt}\right)^{-1}
\left( A_i N_{li}P_i+A_i N_{ri}P_i{N_{ri}}^{-1}N_{li}\right),
\label{eq:ic2}
\end{eqnarray}
\end{widetext}
where $A_t$, $B_t$, and $N_{lt}$ are the values of $A$, $B$, and $N_l$ in the transmitted region obtained by setting $\epsilon=\epsilon_t$ in Eqs.~(\ref{eq:abc}) and (\ref{eq:eta}).
The matrices $P_i$ and $P_t$ in Eqs.~(\ref{eq:imb2}) and (\ref{eq:ic2}) are defined by
\begin{eqnarray}
P_i=\begin{pmatrix} p_{ai} & 0 \\ 0 & p_{bi} \end{pmatrix},~~P_t=\begin{pmatrix} p_{at} & 0 \\ 0 & p_{bt} \end{pmatrix},
\end{eqnarray}
where
\begin{eqnarray}
&&p_{ai}=\sqrt{{k_0}^2{\epsilon_i}^2-{k_y}^2},~~p_{bi}=\sqrt{\frac{1}{4}{k_0}^2{\epsilon_i}^2-{k_y}^2},\nonumber\\&&
p_{at}=\sqrt{{k_0}^2{\epsilon_t}^2-{k_y}^2},~~p_{bt}=\sqrt{\frac{1}{4}{k_0}^2{\epsilon_t}^2-{k_y}^2}.
\label{eq:wns}
\end{eqnarray}
These initial conditions have been obtained following the procedure and using Eq.~(18) given in \cite{sk1}.
If the incident and transmitted regions have the same potential, they reduce to very simple $2\times 2$ matrices $r(0)=0$ and $t(0)=I$.

We assume that the incident waves are propagating waves with a real-valued wave vector. The effective refractive index associated with the $a$ mode is twice as large as that of the $b$ mode. When an $a$ mode wave is incident from the region where $\epsilon_i=1$, $p_{ai}$ is
real and the incident angle $\theta$ is related to $k_y$ by $k_y=k_0\sin\theta$. In this case, if $\theta$ is greater than $30^\circ$, we note that $p_{bi}$ becomes imaginary
and the reflected $b$ wave is evanescent, while the reflected $a$ wave is propagative. However, when a $b$ mode wave is incident, the incident angle $\theta$ is related to $k_y$ by $k_y=(k_0\sin\theta)/2$. Then $p_{ai}$ is always real regardless of the incident angle and both reflected $a$ and $b$ waves are propagative.
The other quantities
$p_{at}$ and $p_{bt}$ can be either real or imaginary depending on the value of $U_t$. For instance, when $p_{at}$ is imaginary, we use $p_{at}=i\sqrt{{k_y}^2-{k_0}^2{\epsilon_t}^2}$ instead of
the expression in Eq.~(\ref{eq:wns}). A sketch of the configuration considered in this section is shown in Fig.~\ref{fig_cf}.

Finally, from the consideration of the probability currents, we obtain the expressions for the reflectance and transmittance matrices $R_{ij}$ and $T_{ij}$ ($i,j=1,2$),
which are applicable when $p_{bi}$, $p_{at}$, and $p_{bt}$ are real:
\begin{eqnarray}
&&R_{11}=\vert r_{11}\vert^2,~~R_{22}=\vert r_{22}\vert^2,\nonumber\\&&
R_{21}=\frac{4p_{bi}}{p_{ai}}\vert r_{21}\vert^2,~~R_{12}=\frac{p_{ai}}{4p_{bi}}\vert r_{12}\vert^2,\nonumber\\
&&T_{11}=\frac{\vert\epsilon_i\vert p_{at}}{\vert\epsilon_t\vert p_{ai}}\vert t_{11}\vert^2,~~T_{22}=\frac{\vert\epsilon_i\vert p_{bt}}{\vert\epsilon_t\vert p_{bi}}\vert t_{22}\vert^2,\nonumber\\&&
T_{21}=\frac{4\vert\epsilon_i\vert p_{bt}}{\vert\epsilon_t\vert p_{ai}}\vert t_{21}\vert^2,~~T_{12}=\frac{\vert\epsilon_i\vert p_{at}}{4\vert\epsilon_t\vert p_{bi}}\vert t_{12}\vert^2.
\end{eqnarray}
If $p_{at}$ ($p_{bt}$) is imaginary, we have to set $T_{11}$ and $T_{12}$ ($T_{21}$ and $T_{22}$) to be identically zero.
If $p_{bi}$ is imaginary, we have to set $R_{21}$ to zero and the mode conversion coefficient $A_2$ (see below) is undefined.
With these definitions, if there is no dissipation or mode conversion, the law of energy conservation $R_{11}+R_{21}+T_{11}+T_{21}=R_{12}+R_{22}+T_{12}+T_{22}=1$ should
be satisfied. When the mode conversion occurs,
the mode conversion coefficients $A_1$ and $A_2$ are defined by
\begin{eqnarray}
&&A_1=1-R_{11}-R_{21}-T_{11}-T_{21},\nonumber\\&& A_2=1-R_{12}-R_{22}-T_{12}-T_{22}.
\end{eqnarray}

In Fig.~\ref{fig3}, we plot the mode conversion coefficients $A_1$ and $A_2$ versus $\theta$, when $\zeta=15$ and $E/U_0=0.3$, 0.6, 0.9. $A_1$ ($A_2$) is obtained by calculating the absorptance when the incident wave is $a$ ($b$) mode.
We find that there exists a wide range of the incident angle in both $A_1$ and $A_2$ curves in which the mode conversion is substantially strong.
Since there are two propagating modes interacting
with the local flat band mode, these curves display multiple peaks and cusps associated with various cutoffs.
Inside the inhomogeneous layer, the $a$ and $b$ modes are coupled to each other. When an $a$ mode wave is incident, it can propagate directly to the resonance region and
convert to flat-band state or it can take an indirect route, first converting to $b$ wave and then converting to flat-band state at the resonance region.
Since the mode conversion coefficient obtains the maximum at different parameter values in these two cases, the curves
of $A_1$ often show two peaks as a function of the incident angle or the energy, as illustrated in Fig.~\ref{fig3}(a).
The case where a $b$ mode wave is incident is significantly different.
Since the refractive index associated with the $a$ mode is twice as large as that of the $b$ mode, the $a$ wave converted from the incident $b$ wave propagates
at a much smaller angle with respect to the $x$ axis than the incident angle. Mode conversion is not efficient for small propagation angles and therefore this
indirect process does not contribute greatly to mode conversion. Therefore, when a $b$ wave is incident, mode conversion is dominated by the direct conversion from $b$ wave to flat-band state and one usually obtains a single peak for $A_2$ as shown in Fig.~\ref{fig3}(b).
In addition, the occurrence of various cutoffs makes cusps to appear in the curves. For example, the cusp at $\theta=30^\circ$ in Fig.~\ref{fig3}(a) comes from the
cutoff condition that the reflected $b$ wave becomes evanescent.

In Fig.~\ref{f1}, we show color graphs of $A_1$ and $A_2$ as a function of the incident angle and the particle energy when $\zeta=15$.
When the energy satisfies $0<E/U_0<1$, there appear wide ranges of the incident angle in which the mode conversion is substantially strong in both graphs.
The observation that there are two clusters in which the mode conversion is strong in Fig.~\ref{f1}(a) has the same reason as that explained in the previous paragraph.

\begin{figure}
\includegraphics[width=\linewidth]{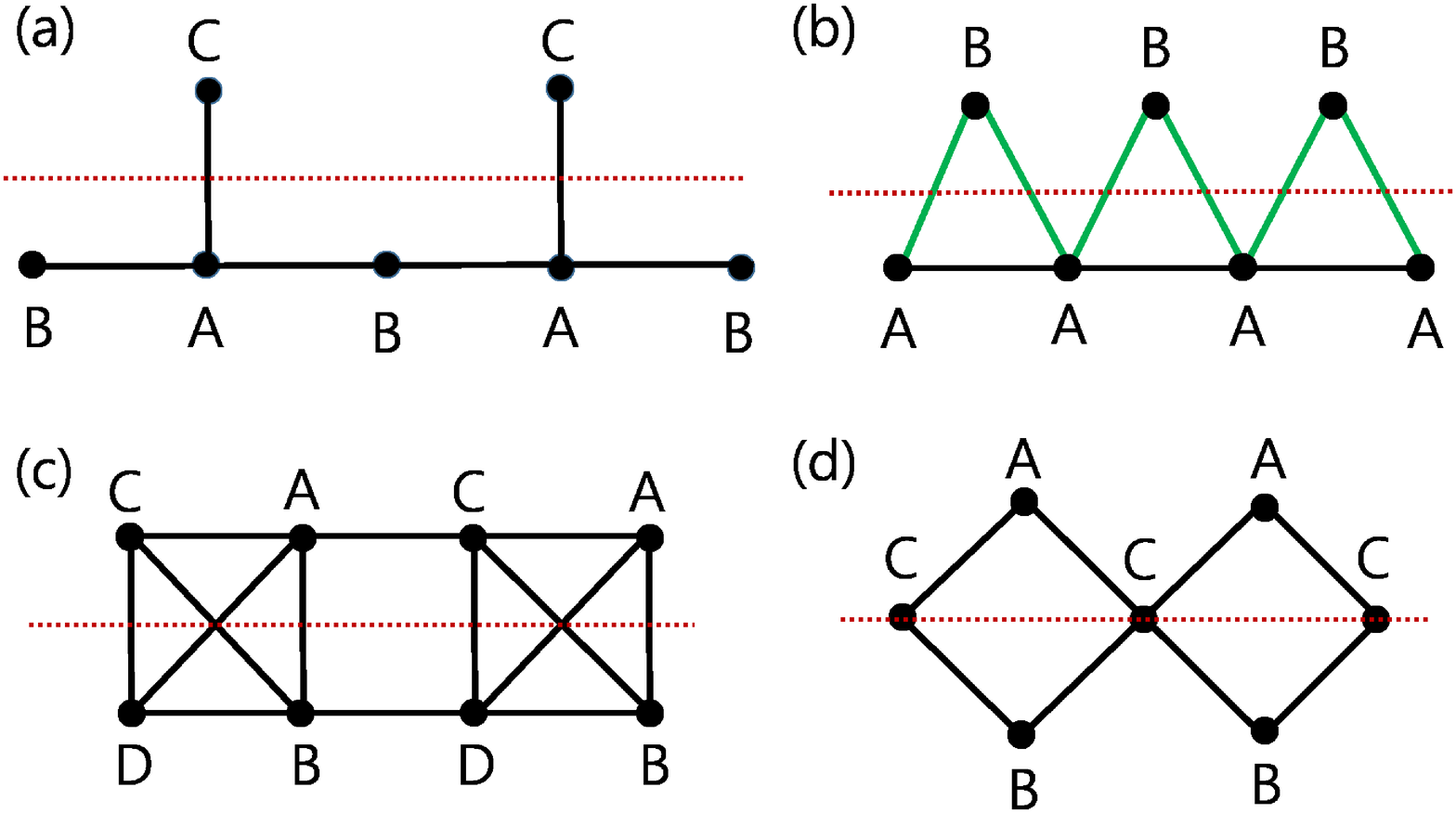}
\caption{Sketches of the (a) stub, (b) sawtooth, (c) 1D pyrochlore, and (d) diamond lattices. In (a), (c), and (d), all hopping parameters between the connected sites are the same. In (b), the hopping parameters associated with the black and green lines are $\tau$ and $\sqrt{2}\tau$, respectively. The 1D pyrochlore and diamond lattices have up-down inversion symmetry, or equivalently, mirror symmetry with respect to the dotted line,
whereas the stub and sawtooth lattices have no such symmetry. In one dimension, mode conversion occurs only in the models derived from the lattices with no up-down inversion symmetry.}
\label{fig4}
\end{figure}

\begin{figure}
\centering\includegraphics[width=8cm]{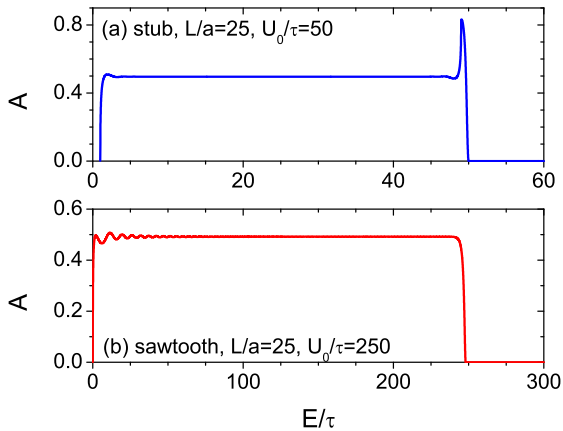}
\caption{Mode conversion coefficient $A$ obtained by solving (a) Eq.~(\ref{eq:stub}) corresponding to the stub lattice when $L/a=25$ and $U_0/\tau=50$ and (b) Eq.~(\ref{eq:saw}) corresponding to the sawtooth lattice when $L/a=25$ and $U_0/\tau=250$ plotted versus normalized energy $E/\tau$. }
\label{fig5}
\end{figure}

\section{1D models with flat bands}

There have been many studies on the lattice models exhibiting one or more flat bands in 1D, including the stub, sawtooth, diamond, 1D pyrochlore, and 1D Lieb lattices \cite{ley,luck,flach}.
We have derived continuum wave equations from these models by linearizing them in the vicinity of the band edge
and solved the wave equations in the presence of an inhomogeneous scalar potential.
All models contain one or more flat bands in addition to dispersive bands.
In order for an incident wave to induce mode conversion, it has to couple to the local flat band mode.
We have found that this coupling is possible only in the models with no up-down inversion symmetry such as the stub and sawtooth lattices (shown in Fig.~\ref{fig4}),
but is not possible in the symmetric models such as the diamond and 1D pyrochlore lattices. These latter cases have a close similarity to the absence of mode conversion at normal incidence in the pseudospin-1
Dirac equation.

The tight-binding equations for the stub lattice at energy $E$ can be written as
\begin{eqnarray}
&& E \psi_n^{\rm A}= v_n^{\rm A}\psi_n^{\rm A}+\tau\psi_{n-1}^{\rm B}+\tau\psi_n^{\rm B}+\tau\psi_n^{\rm C},    \nonumber\\
&& E \psi_n^{\rm B}= v_n^{\rm B}\psi_n^{\rm B}+\tau\psi_{n}^{\rm A}+\tau\psi_{n+1}^{\rm A},    \nonumber\\
&& E \psi_n^{\rm C}= v_n^{\rm C}\psi_n^{\rm C}+\tau\psi_{n}^{\rm A},
\end{eqnarray}
where A, B, and C indicate the sublattice sites A, B, and C shown in Fig.~\ref{fig4}(a) and $v_n^{\rm A}$, $v_n^{\rm B}$, and $v_n^{\rm C}$ are the potentials at each site.
All hopping integrals between the nearest-neighbor sites are assumed to have the same value $\tau$.
In the homogeneous case with no potential, the spectrum of this model is given by the eigenvalues of the matrix
\begin{equation}
\begin{pmatrix} 0 & 1+e^{-iqa} & 1 \\ 1+e^{iqa} &0 &0 \\ 1 & 0 & 0 \end{pmatrix},
\end{equation}
where $q$ is the wave vector and $a$ is the lattice constant.
We find that the spectrum consists of one flat band at $E=0$ and two dispersive bands satisfying $E/\tau=\pm\sqrt{3+2\cos(qa)}$. We linearize this model
in the neighborhood of the band edge, where $q\approx \pi/a$. Finally, we replace $\tilde q$ ($=q-\pi/a$) with the operator $-id/dx$
and reintroduce the scalar potential $U(x)$. We obtain
\begin{eqnarray}
&& E \psi_{\rm A}=U\psi_{\rm A}+\tau a\frac{d\psi_{\rm B}}{dx}+\tau\psi_{\rm C},    \nonumber\\
&& E \psi_{\rm B}=U\psi_{\rm B}-\tau a\frac{d\psi_{\rm A}}{dx},    \nonumber\\
&& E \psi_{\rm C}=U\psi_{\rm C}+\tau\psi_{\rm A}.
\end{eqnarray}
We eliminate $\psi_{\rm A}$ and $\psi_{\rm C}$ and obtain a wave equation of the form
\begin{eqnarray}
a^2\frac{d}{dx}\left(\frac{\varepsilon}{\varepsilon^2-1}\frac{d{\psi_B}}{dx}\right)+\varepsilon\psi_B=0,
\label{eq:stub}
\end{eqnarray}
where $a$ is the lattice constant and
$\varepsilon=[E-U(x)]/\tau$. $\tau$ is the hopping parameter between the neighboring sites and $\psi_{B}$ describes the wave function at the B sites in Fig.~\ref{fig4}(a).
We apply the invariant imbedding method to the wave equation and derive the equations for the reflection and transmission coefficients of the form
\begin{eqnarray}
&&\frac{dr}{dl}=2ip\frac{\varepsilon_1\left(\varepsilon^2-1\right)}{\varepsilon\left({\varepsilon_1}^2-1\right)}r
+\frac{ip}{2}\left[\frac{\varepsilon}{\varepsilon_1}-\frac{\varepsilon_1\left(\varepsilon^2-1\right)}{\varepsilon\left({\varepsilon_1}^2-1\right)}\right]\left(1+r\right)^2,\nonumber\\
&&\frac{dt}{dl}=ip\frac{\varepsilon_1\left(\varepsilon^2-1\right)}{\varepsilon\left({\varepsilon_1}^2-1\right)}t
+\frac{ip}{2}\left[\frac{\varepsilon}{\varepsilon_1}-\frac{\varepsilon_1\left(\varepsilon^2-1\right)}{\varepsilon\left({\varepsilon_1}^2-1\right)}\right]\left(1+r\right)t,\nonumber\\
\label{eq:iest}
\end{eqnarray}
where $\varepsilon_1$ [$=(E-U_1)/\tau=E/\tau$] is the value of $\varepsilon$ in the incident region and $p$ ($=\sqrt{{\varepsilon_1}^2-1}/a$) is the wave number of the incident wave.
These equations are integrated using the initial conditions
\begin{eqnarray}
&&r(0)=\frac{p\varepsilon_1\left({\varepsilon_2}^2-1\right)-p^\prime\varepsilon_2\left({\varepsilon_1}^2-1\right)}
{p\varepsilon_1\left({\varepsilon_2}^2-1\right)+p^\prime\varepsilon_2\left({\varepsilon_1}^2-1\right)},\nonumber\\
&&t(0)=\frac{2p\varepsilon_1\left({\varepsilon_2}^2-1\right)}
{p\varepsilon_1\left({\varepsilon_2}^2-1\right)+p^\prime\varepsilon_2\left({\varepsilon_1}^2-1\right)},
\end{eqnarray}
where $\varepsilon_2$ [$=(E-U_2)/\tau$] is the value of $\varepsilon$ in the transmitted region and $p^\prime$ ($=\sqrt{{\varepsilon_2}^2-1}/a$) is the wave number of the transmitted wave. The reflectance $R$ and the transmittance $T$ are given by
\begin{eqnarray}
R=\vert r\vert^2,~~T=\left\{\begin{matrix} \frac{\vert\varepsilon_2\vert\sqrt{{\varepsilon_1}^2-1}}{\vert\varepsilon_1\vert\sqrt{{\varepsilon_2}^2-1}}\vert t\vert^2
&\mbox{if }{\varepsilon_2}^2>1 \\ 0 & \mbox{if }{\varepsilon_2}^2\le 1 \end{matrix}\right..
\end{eqnarray}
We notice that the invariant imbedding equations, Eq.~(\ref{eq:iest}), have
a singularity at $\varepsilon=0$, which corresponds to $E=U$.

The tight-binding equations for the sawtooth lattice at energy $E$ are written as
\begin{eqnarray}
&& E \psi_n^{\rm A}= v_n^{\rm A}\psi_n^{\rm A}+\tau\psi_{n-1}^{\rm A}+\tau\psi_{n+1}^{\rm A}+\tau^\prime\psi_{n-1}^{\rm B}+\tau^\prime\psi_{n}^{\rm B}, \nonumber\\
&& E \psi_n^{\rm B}= v_n^{\rm B}\psi_n^{\rm B}+\tau^\prime\psi_{n}^{\rm A}+\tau^\prime\psi_{n+1}^{\rm A},
\end{eqnarray}
where A and B indicate the sublattice sites A and B shown in Fig.~\ref{fig4}(b) and $v_n^{\rm A}$ and $v_n^{\rm B}$ are the potentials at each site.
In order for this model to have a flat band, the hopping integral $\tau^\prime$ should be fine-tuned to satisfy $\tau^\prime=\sqrt{2}\tau$.
In the homogeneous case with no potential, the spectrum of this model consists of a flat band at $E/\tau=-2$ and a dispersive band
$E/\tau=2[1+\cos(qa)]$. Following a similar procedure as that for the stub lattice, it is straightforward to derive
\begin{eqnarray}
a^2\frac{d}{dx}\left(\frac{\varepsilon+2}{\varepsilon}\frac{d{\psi_A}}{dx}\right)+(\varepsilon+2)\psi_A=0,
\label{eq:saw}
\end{eqnarray}
where $\psi_A$ describes the wave function at the A sites in Fig.~\ref{fig4}(b).
We apply the invariant imbedding method to it and obtain the equations for the reflection and transmission coefficients:
\begin{eqnarray}
&&\frac{dr}{dl}=2ip\frac{\varepsilon\left(\varepsilon_1+2\right)}{\varepsilon_1\left(\varepsilon+2\right)}r
+\frac{ip}{2}\left[\frac{\varepsilon+2}{\varepsilon_1+2}-\frac{\varepsilon\left(\varepsilon_1+2\right)}{\varepsilon_1\left(\varepsilon+2\right)}\right]\left(1+r\right)^2,\nonumber\\
&&\frac{dt}{dl}=ip\frac{\varepsilon\left(\varepsilon_1+2\right)}{\varepsilon_1\left(\varepsilon+2\right)}t
+\frac{ip}{2}\left[\frac{\varepsilon+2}{\varepsilon_1+2}-\frac{\varepsilon\left(\varepsilon_1+2\right)}{\varepsilon_1\left(\varepsilon+2\right)}\right]\left(1+r\right)t,\nonumber\\
\label{eq:iesaw}
\end{eqnarray}
where $\varepsilon_1$ [$=(E-U_1)/\tau=E/\tau$] is the value of $\varepsilon$ in the incident region and $p$ ($=\sqrt{\varepsilon_1}/a$) is the wave number of the incident wave.
These equations are integrated using the initial conditions
\begin{eqnarray}
&&r(0)=\frac{p\varepsilon_2\left(\varepsilon_1+2\right)-p^\prime\varepsilon_1\left(\varepsilon_2+2\right)}
{p\varepsilon_2\left(\varepsilon_1+2\right)+p^\prime\varepsilon_1\left(\varepsilon_2+2\right)},\nonumber\\
&&t(0)=\frac{2p\varepsilon_2\left(\varepsilon_1+2\right)}
{p\varepsilon_2\left(\varepsilon_1+2\right)+p^\prime\varepsilon_1\left(\varepsilon_2+2\right)},
\end{eqnarray}
where $\varepsilon_2$ [$=(E-U_2)/\tau$] is the value of $\varepsilon$ in the transmitted region and $p^\prime$ ($=\sqrt{\varepsilon_2}/a$) is the wave number of the transmitted wave. The reflectance $R$ and the transmittance $T$ are given by
\begin{eqnarray}
R=\vert r\vert^2,~~T=\left\{\begin{matrix} \frac{\sqrt{\varepsilon_1}\left(\varepsilon_2+2\right)}{\sqrt{\varepsilon_2}\left(\varepsilon_1+2\right)}\vert t\vert^2
&\mbox{if }\varepsilon_2> 0 \\ 0 & \mbox{if }\varepsilon_2\le 0 \end{matrix}\right..
\end{eqnarray}
We notice that the invariant imbedding equations, Eq.~(\ref{eq:iesaw}), have
a singularity at
$\varepsilon=-2$, that is, $E=U-2\tau$.

In Fig.~\ref{fig5}(a), we plot the
mode conversion coefficient $A$ obtained by solving Eq.~(\ref{eq:stub}) when $L/a=25$ and $U_0/\tau=50$ versus normalized energy $E/\tau$.
In Fig.~\ref{fig5}(b), we plot $A$ obtained by solving Eq.~(\ref{eq:saw})
when $L/a=25$ and $U_0/\tau=250$. The configuration of the potential is given by Eq.~(\ref{eq:slow})
in both cases. In Fig.~\ref{fig5}(a), $A$ is nonzero in the range $1<E/\tau<U_0/\tau ~(=50)$, since the wave number of the incident wave is real when $E/\tau>1$
and the resonance point exists when $0<E/\tau<U_0/\tau$. When $E/\tau>49$, the wave number in the transmitted region becomes imaginary and the wave gets strongly reflected, which
causes a sharp peak to occur in the region $49<E/\tau<50$. Except for this region, $A$ is almost a constant approximately equal to 0.5.
In Fig.~\ref{fig5}(b), $A$ is nonzero in the range $0<E/\tau<248$, since the resonance point exists when $0<E/\tau<(U_0/\tau)-2$. $A$ is approximately equal to 0.49 in the entire nonzero range. We also remind that the continuum models used in our calculation have been derived using the assumption that the wavelength, which is inversely proportional to the energy, is sufficiently larger than the lattice constant $a$. Therefore we expect that our results are quantitatively valid in the low energy region.

\begin{figure}
\centering\includegraphics[width=8cm]{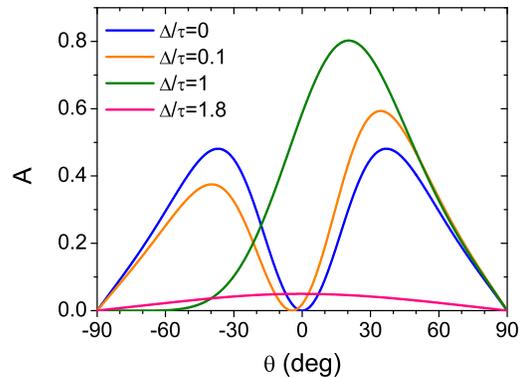}
\caption{Mode conversion coefficient $A$ for the 2D model described by Eq.~(\ref{eq:sm}) in the configuration given by Eq.~(\ref{eq:slow}) plotted versus incident angle $\theta$, when $L/a=10$, $U_0/\tau=20$, $E/\tau=2$, and $\Delta/\tau=0$, 0.1, 1, 1.8.}  
\label{fig6}
\end{figure}

\section{2D model with a nearly flat band}

Mode conversion can also occur when a dispersive band coexists with a nearly flat band with
a small group velocity. An example in plasma physics is the conversion of a transverse EM wave into an electron plasma wave
in a warm plasma at nonzero temperature \cite{yu1}. Here we consider the 2D model with a singular flat band studied in \cite{im2}, the Hamiltonian for which is given by
\begin{eqnarray}
{\mathcal H}=\begin{pmatrix} \tau a^2 {k_x}^2 +U(x) & \Delta-i \tau a^2 k_x k_y\\ \Delta+i\tau a^2 k_x k_y & \tau a^2 {k_y}^2 +U(x) \end{pmatrix},
\label{eq:sm}
\end{eqnarray}
where $\tau$ and $a$ are the energy and length scales of the model and $k_x$ ($=-id/dx$) is a differential operator.
When the parameter $\Delta$ and the potential $U$ are zero, the model has a quadratic dispersive band touching a flat band with $E=0$ at $k_x=k_y=0$.
As $\Delta$ increases from zero, a gap is opened and
the flat band becomes increasingly warped.

When the potential $U$ is uniform, the eigenvalues of Eq.~(\ref{eq:sm}) are given by
\begin{eqnarray}
E=U+\frac{\tau a^2k^2}{2}\pm\sqrt{\left(\frac{\tau a^2k^2}{2}\right)^2+\Delta^2},
\label{eq:im}
\end{eqnarray}
where $k^2={k_x}^2+{k_y}^2$.
When the parameter $\Delta$ is zero, the spectrum corresponding to the minus sign in Eq.~(\ref{eq:im}) is a flat band with $E=U$.
As $\Delta$ increases from zero, this band becomes warped and dispersive, though the group velocity remains small when $\Delta$ is sufficiently small.
We assume that the potential $U$ is a function of $x$ and replace $k_x$ with the operator $k_x=-id/dx$. We obtain
\begin{eqnarray}
&& (E-U)\psi_1=-\tau a^2\frac{d^2\psi_1}{dx^2}+\Delta \psi_2-\tau a^2 k_y\frac{d\psi_2}{dx},   \nonumber\\
&& (E-U)\psi_2=\Delta \psi_1+\tau a^2 k_y\frac{d\psi_1}{dx}+\tau a^2 {k_y}^2\psi_2,
\end{eqnarray}
from which we derive the invariant imbedding equations for $r$ and $t$ of the form
\begin{widetext}
\begin{eqnarray}
&&\frac{dr}{dl}=-\frac{\Delta }{E-U}k_y(r+1)+\frac{E-U-\tau a^2{k_y}^2}{E-U}\left[\frac{E}{E-\tau a^2{k_y}^2}ip(r-1)+\frac{\Delta }{E-\tau a^2{k_y}^2}k_y(r+1)\right]\nonumber\\&&~~~~~~+\left[\left(ip+\frac{\Delta }{E}k_y\right)\frac{\Delta }{E-U}k_y
-\frac{E-U}{\tau a^2}\frac{E-\tau a^2{k_y}^2}{E}+\frac{1}{\tau a^2}\frac{\Delta^2}{E-U}\frac{E-\tau a^2{k_y}^2}{E}\right]\frac{1}{2ip}(r+1)^2\nonumber\\
&&~~~~~~+\left[-\left(ip+\frac{\Delta }{E}k_y\right)\frac{E-U-\tau a^2{k_y}^2}{E-U}+\frac{\Delta }{E-U}\frac{E-\tau a^2{k_y}^2}{E}k_y\right]\frac{1}{2ip}(r+1)
\nonumber\\&&~~~~~~~~~\times\left[\frac{E}{E-\tau a^2{k_y}^2}ip(r-1)+\frac{\Delta }{E-\tau a^2{k_y}^2}k_y(r+1)\right],\nonumber\\
&&\frac{dt}{dl}=\left[\left(ip+\frac{\Delta }{E}k_y\right)\frac{\Delta }{E-U}k_y
-\frac{E-U}{\tau a^2}\frac{E-\tau a^2{k_y}^2}{E}+\frac{1}{\tau a^2}\frac{\Delta^2}{E-U}\frac{E-\tau a^2{k_y}^2}{E}\right]\frac{1}{2ip}t(r+1)\nonumber\\
&&~~~~~~+\left[-\left(ip+\frac{\Delta }{E}k_y\right)\frac{E-U-\tau a^2{k_y}^2}{E-U}+\frac{\Delta }{E-U}\frac{E-\tau a^2{k_y}^2}{E}k_y\right]\frac{1}{2ip}t
\nonumber\\&&~~~~~~~~~\times\left[\frac{E}{E-\tau a^2{k_y}^2}ip(r-1)+\frac{\Delta }{E-\tau a^2{k_y}^2}k_y(r+1)\right],
\label{eq:im2}
\end{eqnarray}
\end{widetext}
where we have assumed that the potential $U$ in the incident region vanishes and the wave number of the incident wave, $p$, is given by
\begin{eqnarray}
p=\sqrt{\frac{E(E-\tau a^2{k_y}^2)-\Delta^2}{\tau a^2 E}}.
\end{eqnarray}
These equations are integrated using the initial conditions
\begin{widetext}
\begin{eqnarray}
&&r(0)=\frac{\left(ip-\frac{\Delta }{E}k_y\right)\frac{E-U_t-\tau a^2 {k_y}^2}{E-U_t}+\left(-ip^\prime+\frac{\Delta }{E-U_t}k_y\right)\frac{E-\tau a^2 {k_y}^2}{E}}{\left(ip+\frac{\Delta }{E}k_y\right)\frac{E-U_t-\tau a^2{k_y}^2}{E-U_t}
+\left(ip^\prime-\frac{\Delta }{E-U_t}k_y\right)\frac{E-\tau a^2{k_y}^2}{E}},\nonumber\\
&&t(0)=\frac{2ip\frac{E-U_t-\tau a^2{k_y}^2}{E-U_t}}{\left(ip+\frac{\Delta }{E}k_y\right)\frac{E-U_t-\tau a^2{k_y}^2}{E-U_t}+\left(ip^\prime-\frac{\Delta }{E-U_t}k_y\right)\frac{E-\tau a^2{k_y}^2}{E}},
\end{eqnarray}
\end{widetext}
where $U_t$ is the value of $U$ in the transmitted region and the wave number of the transmitted wave, $p^\prime$, is given by
\begin{eqnarray}
p^\prime=\sqrt{\frac{(E-U_t)(E-U_t-\tau a^2{k_y}^2)-\Delta^2}{\tau a^2 (E-U_t)}}.
\end{eqnarray}
The reflectance $R$ and the transmittance $T$ are obtained using
\begin{eqnarray}
&&R=\vert r\vert^2,\nonumber\\ &&T=\left\{\begin{matrix} \frac{p^\prime}{p}\bigg\vert\frac{(E-\tau a^2{k_y}^2)(E-U_t)}{(E-U_t-\tau a^2{k_y}^2)E}\bigg\vert\vert t\vert^2
&\mbox{if $p^\prime$ is real}  \\ 0 & \mbox{if $p^\prime$ is imaginary} \end{matrix}\right..\nonumber\\
\end{eqnarray}
We notice that the invariant imbedding equations, Eq.~(\ref{eq:im2}), have
a singularity at
$E=U$.

In Fig.~\ref{fig6}, we plot the mode conversion coefficient versus incident angle, when the potential is given by Eq.~(\ref{eq:slow}) and the parameters are
$L/a=10$, $U_0/\tau=20$, $E/\tau=2$, and $\Delta/\tau=0$, 0.1, 1, 1.8. When $\Delta=0$, $A$ is symmetric under the sign change of $\theta$.
As $\Delta/\tau$ increases towards 1, $A$ becomes increasingly asymmetric. As $\Delta/\tau$ increases above 1,
the mode conversion becomes less and less efficient and $A$ decreases rapidly towards zero.
We find that the overall behavior is quite similar to the mode conversion of extraordinary waves occurring in a magnetized plasma where the external magnetic field
is perpendicular to the directions of inhomogeneity and wave propagation \cite{jkps}. The parameter $\Delta$ plays the role of the magnetic field strength.
We conclude that the mode conversion between a strongly dispersive band and a nearly flat band can occur, but its efficiency decreases rapidly as the nearly flat band becomes less and less flat.

\section{Conclusion}

In this paper,
we have demonstrated that mode conversion and resonant absorption occur generically in all systems where dispersive bands and flat bands coexist
in the band structure, unless forbidden by symmetry.
Mode conversion takes place as long as there exists a resonant region inside the system, regardless of the shape of
the potential.
In electronic materials, the inhomogeneous potential can be easily achieved by applying an external electric field or strain.
A slowly-varying inhomogeneity in a medium parameter can be introduced into photonic metamaterials by various methods,
for instance, by slowly varying the lattice constant in a certain direction.
The effects of mode conversion in metamaterials are manifested in many properties, most obviously in the absorption of the incident wave.
In electronic systems, the mode conversion effect will substantially influence many physical quantities including the conductance and the shot noise.
We will present a detailed analysis on the consequences of mode conversion in experimentally relevant systems in a future work.

\acknowledgments
This research was supported through a National Research
Foundation of Korea Grant (NRF-2020R1A2C1007655)
funded by the Korean Government.
It was also supported by the Basic Science Research Program funded by the Ministry of Education (2021R1A6A1A10044950) and by
the Global Frontier Program (2014M3A6B3063708).

\end{document}